\documentclass[a4paper,fleqn,usenatbib]{mnras}

\usepackage{newtxtext,newtxmath}

\usepackage[T1]{fontenc}
\usepackage{ae,aecompl}

\usepackage{graphicx}	
\usepackage{amsmath}	
\usepackage{pdflscape}	
\usepackage{multicol}
\usepackage{afterpage}
\usepackage{color}
\usepackage{etoolbox}
\usepackage{booktabs}
\usepackage[utf8]{inputenc}
\usepackage{ulem}

\definecolor{darkmagenta}{rgb}{0.5, 0, 0.5}
\definecolor{darkgreen}{rgb}{0, 0.6, 0.05}
\definecolor{darkred}{rgb}{0.86,0.078,0.235}

\title[SDSS J1004+4112]{Free-Form and Hybrid Lens Models for SDSS J1004+4112: Substructure and Central Image Time Delay Constraints}

\author[Derek Perera et al.]{Derek Perera$^{1}$\thanks{E-mail: perer030@umn.edu},
Liliya L. R. Williams$^{1}$, Jori Liesenborgs$^{2}$, Agniva Ghosh$^{1}$, Prasenjit Saha$^{3}$
\\
$^{1}$School of Physics and Astronomy, University of Minnesota, Minneapolis, MN, 55455, USA.\\
$^{2}$UHasselt – Flanders Make, Expertisecentrum voor Digitale Media, Wetenschapspark 2, B-3590, Diepenbeek, Belgium.\\
$^{3}$Physik-Institut, University of Zurich, Winterthurerstrasse 190, 8057 Zurich, Switzerland.\\
}
\date{Accepted XXX. Received YYY; in original form ZZZ}

\pubyear{2023}

\begin{document}
\label{firstpage}
\pagerange{\pageref{firstpage}--\pageref{lastpage}}
\maketitle

\begin{abstract}
    SDSS J1004+4112 is a well studied gravitational lens with a recently measured time delay between its first and fourth arriving quasar images. Using this new constraint, we present updated free-form lens reconstructions using the lens inversion method {\tt GRALE}, which only uses multiple image and time delay data as inputs. In addition, we obtain hybrid lens reconstructions by including a model of the brightest cluster galaxy (BCG) as a Sersic lens. For both reconstructions, we use two sets of images as input: one with all identified images, and the other a revised set leaving out images that have been potentially misidentified. We also develop a source position optimization MCMC routine, performed on completed {\tt GRALE} runs, that allows each model to better match observed image positions and time delays. All the reconstructions produce similar mass distributions, with the hybrid models finding a steeper profile in the center. Similarly, all the mass distributions are fit by the Navarro-Frenk-White (NFW) profile, finding results consistent with previous parametric reconstructions and those derived from Chandra X-ray observations. We identify a $\sim 5 \times 10^{11} M_{\odot}$ substructure apparently unaffiliated with any cluster member galaxy and present in all our models, and study its reality. Using our free-form and hybrid models we predict a central quasar image time delay of $\sim 2980 \pm 270$ and $\sim 3280 \pm 215$ days, respectively. A potential future measurement of this time delay will, while being an observational challenge, further constrain the steepness of the central density profile.
\end{abstract}



\section{Introduction}

Galaxy clusters, despite containing hundreds to thousands of individual galaxies and a hot intracluster plasma, are dominated almost entirely by dark matter. Because of this, observations are not sufficient to completely model the mass distributions of galaxy clusters. One way around this is to utilize gravitational lensing to map the mass distribution. In such scenarios, the galaxy cluster lenses the background source(s) into multiple point images and/or arcs, whose positions and magnifications are entirely dependent on the mass distribution of the cluster.

One such well known example is SDSS J1004+4112. In this lens, the cluster located at $z_d = 0.68$ lenses a quasar ($z = 1.734$) into a "quad" configuration, with 4 bright outer images with maximum separation of 14.62 arcseconds \citep{inada03}, and a spectroscopically confirmed demagnified central image \citep{inada05,inada08}. Time delays have been measured for all the quasar images except the central image \citep{fohlmeister07,fohlmeister08,munoz22}. In addition, there are at least 3 other galaxy sources in the source plane \citep{sharon05}. SDSS J1004+4112 has been observed and studied extensively in X-ray \citep{ota06,lamer06,chen12}, visual and infrared \citep{ross09}, and radio \citep{jackson11,mckean21,hartley21} wavelengths. Microlensing variability of the quasar images has been observed using these methods and have helped study the quasar accretion disk size \citep{fian16}, quasar image flux ratio \citep{lamer06}, and broad emission lines \citep{gomezalvarez06,fian18,popovic20,hutsemekers23}.

What makes SDSS J1004+4112 an object of considerable interest are the precisely measured time delays, and the presence of a central image in the quasar source. The central image is typically strongly demagnified beyond visibility, especially in individual galaxy lenses or when they are located very close to a bright cluster galaxy. It has been suggested that central images in quads can be detected \citep{hezaveh15,quinn16,perera23}. These observational constraints uniquely allow one to study in detail the central mass distribution of SDSS J1004+4112. Particularly, lens modelling using these observations can place constraints on dark matter substructures, which in turn provide constraints on the nature of dark matter, such as its interaction cross section \citep{kneib11,peter13} or through its substructure mass fraction \citep{lagattuta23,oriordan23}. Time delay predictions from sophisticated mass models of galaxy clusters can be used to obtain independent and more precise measurements of the Hubble constant $H_0$ \citep{birrer22,napier23,kelly23,liu23}.

SDSS J1004+4112 has been modelled using both parametric \citep{inada03,sharon05,fohlmeister07,oguri10,forestoribio22,napier23,liu23} and non-parametric (free-form) \citep{williams04,liesenborgs09,mohammed15} methods. Parametric models generally model individual cluster member galaxies and the broader dark matter profile with simple mass profiles such as Navorro-Frenk-White (NFW) \citep{nfw97}, or pseudo-Jaffe \citep{keeton01}. While these models are physically motivated by properties of the cluster, their strict priors may be unable to identify all the features of the mass distribution. Free-form models offer a flexible alternative using only observed image positions and/or time delays as input, and can thus offer an unbiased measure of the mass distribution. The drawback of these models is that they are not constrained by existing stellar mass in cluster galaxy members. Hybrid models utilize the physically motivated priors from parametric methods in addition to free-form model inputs, and serve as a middle ground between the parametric and free-form methods.

Here, we use free-form and hybrid models to develop updated mass distribution maps for SDSS J1004+4112. This work is the first free-form lens model for SDSS J1004+4112 that includes the most recently measured time delay \citep{munoz22}. Using these mass distribution maps, we can identify dark matter substructures and compare our result with a NFW profile and previously determined mass profiles. We also use our models to predict the central quasar image time delay.

In Section \ref{txt:data} we present the observed image positions and time delays for SDSS J1004+4112; in Section \ref{txt:recon} we discuss our lens modelling methodology; in Section \ref{txt:results} we present the results from our analysis; in Section \ref{txt:conclusion} we discuss the implications of our results and possibilities for future work. We assume flat $\Lambda$CDM cosmology throughout this paper with $\Omega_M = 0.27$, $\Omega_{\Lambda} = 0.73$, and $H_0 = 70$ km s$^{-1}$ Mpc$^{-1}$.

\section{Data}\label{txt:data}

\begin{figure}
    \centering
    \includegraphics[trim={6.9cm 0.9cm 6.9cm 0.37cm},clip,width=0.49\textwidth]{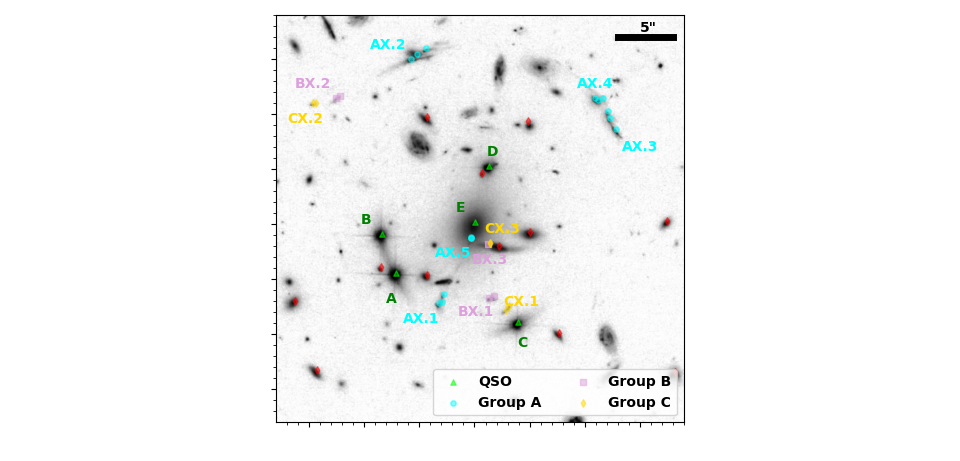}    
\caption{SDSS J1004+4112. Images are presented for the QSO (green), Group A (cyan), Group B (purple), and Group C (gold). For Groups A, B, and C, labels are given using the image names from Table \ref{tab:data}, with X representing the galaxy source within the group. X $\in [1,3]$ for Group A and X $\in [1,2]$ for Groups B and C. Red diamonds indicate identified cluster galaxy members \citep{oguri10}, not including the BCG located near to QSO image E. For the QSO `quad', the arrival sequence of images is apparent from their configuration \citep{sah03}, and is confirmed by the time delay measurements: C, B, A, D, and lastly, E. For the 3 galaxy image Groups, the arrival sequence is also clear from their configuration. For this image, north is up and west is right.}
\label{fig:SDSSJ1004}
\end{figure}

\begin{table}
	\caption{Observational Data}
	\begin{tabular}{lllllc} 
		\hline
		Source & Redshift & Image Name & x [arcsec] & y [arcsec] \\
		\hline
            & & A & 0.000 & 0.000 & \\
            & & B & -1.317 & 3.532 & \\
            QSO & 1.734 & C & 11.039 & -4.492 & \\
            & & D & 8.399 & 9.707 & \\
            & & E & 7.197 & 4.603 & \\
		\hline
            & & A1.1 & 3.93 & -2.78 & \\
            & & A1.2 & 1.33 & 19.37 & \\
            & & A1.3 & 19.23 & 14.67 & \\
            & & A1.4 & 18.83 & 15.87 & \\
            & & A1.5 & 6.83 & 3.22 & \\ \cmidrule{3-5}
            & & A2.1 & 4.13 & -2.68 & \\
            & & A2.2 & 1.93 & 19.87 & \\
            Group A & 3.33 & A2.3 & 19.43 & 14.02 & \\
            & & A2.4 & 18.33 & 15.72 & \\
            & & A2.5 & 6.83 & 3.12 & \\ \cmidrule{3-5}
            & & A3.1 & 4.33 & -1.98 & \\
            & & A3.2 & 2.73 & 20.37 & \\
            & & A3.3 & 19.95 & 13.04 & \\
            & & A3.4 & 18.03 & 15.87 & \\
            & & A3.5 & 6.83 & 3.02 & \\
		\hline
            & & B1.1 & 8.88 & -2.16 & \\
            & & B1.2 & -5.45 & 15.84 & \\
            & & B1.3 & 8.33 & 2.57 & \\ \cmidrule{3-5}
            Group B & 2.74 & B2.1 & 8.45 & -2.26 & \\
            & & B2.2 & -5.07 & 16.04 & \\
            & & B2.3 & 8.33 & 2.57 & \\
		\hline
            & & C1.1 & 10.25 & -3.06 & \\
            & & C1.2 & -7.55 & 15.39 & \\
            & & C1.3 & 8.49 & 2.72 & \\ \cmidrule{3-5}
            Group C & 3.28 & C2.1 & 9.95 & -3.36 & \\
            & & C2.2 & -7.30 & 15.44 & \\
            & & C2.3 & 8.49 & 2.72 & \\
		\hline
	\end{tabular}\\
\medskip{Observed images and classifications for SDSS J1004+4112 as used in \cite{oguri10,forestoribio22}. QSO images have a positional uncertainty of $0.04"$. The images for Groups A, B, and C are assumed to have positional uncertainty of $0.4"$.}

\label{tab:data}
\end{table}

Figure \ref{fig:SDSSJ1004} shows the identified images used in this study, which are also presented in Table \ref{tab:data}. We adopt the image labels used by \cite{oguri10, forestoribio22}. The quasar images are identified by \cite{inada05} with the Hubble Space Telescope Advanced Camera for Surveys (HST/ACS) F814W image, while the image positions for Groups A, B, and C are identified by \cite{sharon05} using HST/ACS F435W, F555W, and F814W images.

At a redshift of 1.734, a quasar (QSO) is lensed by the cluster ($z_d$ = 0.68), forming a quadruple image system with a largest image separation of 14.62 arcseconds \citep{inada03} (QSO A-E). Throughout our analysis, we set QSO image A to be at the center of the coordinate system. \cite{inada05} discovered the fifth central image (QSO E) located $\sim$0.2 arcseconds from the BCG. The arriving order of the QSO images is: C-B-A-D-E. Time delays have been measured relative to the first arriving QSO image (QSO C) for the following (also see Table \ref{tab:tdelayresults}): time delay between B and C $\Delta t_{BC} = 781.92 \pm 2.20$ days \citep{fohlmeister07,fohlmeister08}, time delay between A and C $\Delta t_{AC} = 825.99 \pm 2.10$ days \citep{fohlmeister08}, time delay between D and C $\Delta t_{DC} = 2456.99 \pm 5.55$ days \citep{munoz22}.

In addition to the QSO, there are 3 other multiply imaged sources in this lens \citep{sharon05}: Group A at $z_s = 3.33$, Group B at $z_s = 2.74$, and Group C at $z_s = 3.28$. The image positions for each of these groups have a large uncertainty of $0.4"$ due to uncertainty in locating each images' centroid \citep{oguri10} caused by ambiguous identification of each respective extended image-counterimage pair. Some studies report Groups A, B, and C as a single source galaxy with 5, 3, and 3 images, respectively \citep{sharon05,liesenborgs09,mohammed15}. Others report 3 galaxies in Group A (each with 5 images) and 2 galaxies in both Groups B and C (each with 3 images) \citep{oguri10,forestoribio22}. In this study, we adopt the latter image classifications. Due to the described discrepancy, we simultaneously study models using image positions informed by the former image classifications for comparison.

The brightest cluster galaxy (BCG) is located at (7.114", 4.409"), and was measured to have a velocity dispersion of $352 \pm 13$ km s$^{-1}$ \citep{inada08}. Using this measurement and the relations described in \cite{jorgensen13}, we estimate the dynamical mass $M_{dyn}$ and effective radius $R_e$ of the BCG to be $\sim 10^{12.2} M_{\odot}$ and $\sim 10^{1.2}$ kpc, respectively.

Lastly, we calculate the critical surface density for the QSO source $\Sigma_{\rm crit}$ to be $1.17 \times 10^{11} M_{\odot}$ arcsec$^{-2}$.

\section{Lens Reconstructions}\label{txt:recon}
Our lens modelling procedure involves two sequential steps. First, using {\tt GRALE} (see Section \ref{txt:grale}, we generate 40 individual lens solutions. This gives us a sample of 40 lens models, each with varying fitness values, reconstructed source positions, and QSO time delays. After this, we average over this sample of 40 lens models to get one mean lens model which we use as our best overall lens model. Even for this mean model, the back-projected images will not coincide exactly to well defined source positions. To determine which precise source positions to use, we use the Metropolis-Hastings algorithm to optimize them (see Section \ref{txt:optim}).

\subsection{Lens Reconstruction with Grale}\label{txt:grale}

In this work, we use the non-parametric (free-form) lens inversion software {\tt GRALE} \footnote{{\tt GRALE} is publicly available, and the software can be found at the following page: \url{https://research.edm.uhasselt.be/~jori/grale2/index.html}}. {\tt GRALE} utilizes a grid lens inversion technique based on a genetic algorithm to optimize solutions to the lensing equations \citep{liesenborgs06,liesenborgs07,liesenborgs20}. As a result of this, it is a flexible inversion method, and only requires lens image locations and redshifts as input. In a nutshell, {\tt GRALE} optimizes a mass basis of projected Plummer spheres on a grid. A Plummer sphere has projected surface mass density of:
\begin{equation}\label{eq:plummerdens}
    \Sigma(\theta) = \frac{M}{\pi D_d^2}\frac{\theta_P^2}{(\theta^2 + \theta_P^2)^2},
\end{equation}
and a lens potential of:
\begin{equation}\label{eq:plummerlpot}
    \psi(\theta) = \frac{2GM D_{ds}}{c^2 D_s D_d} \ln{(\theta^2 + \theta_P^2)},
\end{equation}
where $\theta_P$ is the characteristic angular width of the Plummer sphere, $M$ is its total mass, and $D$'s are angular diameter distances between the observer, {\it s}ource and {\it d}eflector. The inversion method seeks to optimize the weights of all the Plummers in the grid, in essence by trying to make backprojected images overlap in the source plane (this is quantified by the "pointimageoverlap" fitness measure described below). After optimizing, the resulting grid is steadily refined, with regions of more mass receiving more refinement. This process continues over many generations. At each generation of the genetic algorithm, weights are optimized based on several fitness criteria. In our case, we use 4 fitness measures (in order of priority):
\begin{itemize}
    \item "pointimagenull": A uniform null space map of triangles is inputted where no images are expected to be formed. This map is back projected into the source plane and penalizes regions where the null space map overlaps with the estimated sources. See \cite{zitrin10} for more information.
    \item "pointgroupoverlap": Specific images of a source are back projected to the source plane. This source position is then forward projected to get the resulting multiple images, which are then used to calculate the RMS of the images in the lens plane. A lower RMS results in a better fitness (lower fitness value). It should be mentioned that the forward projection of the source position is approximated by a one-step procedure, where the difference in the source plane is mapped onto a difference in the image plane using the magnification matrix. See \cite{liesenborgs20} for more details.
    \item "pointimageoverlap": Simply measures how well images of a source overlap when back projected into the source plane. The scale on which the overlap is measured is set by the region of all back projected images, which guards against overfocusing. This consequentially optimizes the source position for each individual {\tt GRALE} run. This is described in \cite{zitrin10} in more detail. 
    \item "timedelay": When time delay information is provided, this fitness criteria is calculated by measuring how well the model reconstructed the observed time delays according to equation 18 of \cite{liesenborgs20}.
\end{itemize}
In general, regions with large mass density will have a more refined grid of Plummers than regions of lower density because as the genetic algorithm optimizes the Plummer weights over many generations, it will subdivide the high density regions further. For a given {\tt GRALE} run, the number of subdivisions is steadily increased after the previous grid has been optimized according to the above fitness criteria. The overall best inversion model is therefore the subdivision grid with the best fitness values. This means that for a number of {\tt GRALE} runs the best fitting mass model can vary in the number of Plummers, usually on the order of a few thousand. The final output is a grid of varying size cells each consisting of a Plummer with mass $M$, determined from the inversion, and width $\theta_P$, proportional to the cell size. We note that we are not limited by the grid resolution since all the components are analytical functions and can be calculated to arbitrary precision at any location in the lens plane.

In addition to Plummers, other basis functions can be added to the initial grid\footnote{While we use Plummers (the default basis function used by {\tt GRALE}), the user is not restricted to this, and can perform an inversion without any Plummers if desired. One can use whatever basis functions that {\tt GRALE} provides.}. These are typically simple parametric lens models\footnote{A full list of lens models incorporated into {\tt GRALE} can be found here: \url{https://research.edm.uhasselt.be/~jori/grale2/grale_lenses.html}} that can assist in modelling specific regions of the cluster (e.g. cluster members). If used, these are added to the basis grid and optimized alongside the Plummers according to the same procedure \citep{liesenborgs20}.

\subsection{Source Position Optimization}\label{txt:optim}
After all individual {\tt GRALE} runs are completed, we use the best fitting model from each {\tt GRALE} run and calculate the surface mass density with equation \ref{eq:plummerdens} for each Plummer. Similarly, with equation \ref{eq:plummerlpot} we calculate the time delay surface of the lens model:
\begin{equation}\label{eq:tdelay}
    \tau(\boldsymbol{\theta}) = \frac{(1+z_d)}{c}\frac{D_d D_s}{D_{ds}}\left(\frac{1}{2}\left(\boldsymbol{\theta}-\boldsymbol{\beta}\right)^2 - \psi(\boldsymbol{\theta})\right)
\end{equation}
where $\boldsymbol{\beta} = (\widetilde{x_s},\widetilde{y_s})$ for any given run. Since each individual {\tt GRALE} run uses a different random seed, each run produces a different lens model of varying fitness and with varying source positions. To account for this, it is best to average over a number of runs and use the mean solution as the best overall lens model. Importantly, the corresponding mean source position $\boldsymbol{\overline{\beta}} = (\overline{x_s},\overline{y_s})$ of the best overall lens model may not be the optimal source position. This best model will not back-project each image to the exact same position in the source plane, this will only be approximately so. In previous papers using {\tt GRALE}, the corresponding mean position of the back-projected images is chosen as the true source position, but this may not be the optimal one. As a consequence, when forward projecting $\boldsymbol{\beta}$ to the lens plane, there will be a natural disagreement between the reconstructed image positions ($x'_i$,$y'_i$) and reconstructed QSO time delays $\Delta t'_{iC}$, and the observations. Therefore, in order to better fit the observations, we seek to optimize $\boldsymbol{\beta}$ for the mean model for all sources. We emphasize that this resulting method (described below) to find the optimized source positions occurs after averaging over a number of {\tt GRALE} runs, and is independent of the "pointimageoverlap" optimization performed during an individual {\tt GRALE} run.

To achieve this, after averaging over all the {\tt GRALE runs}, we use the Metropolis-Hastings algorithm, a MCMC method that will gradually approach a best fit solution based on each iteration's Bayesian posterior. For our purposes, we use the mean source position $\boldsymbol{\overline{\beta}}$ of the averaged lens model as the initial state of the chain. At each iteration, a new test source position $\boldsymbol{\beta}_{n+1}$ is drawn from a Gaussian centered at the previous source position $\boldsymbol{\beta}_{n}$ with a standard deviation of $0.04"$, equivalent to the astrometric accuracy of HST. From here, we compare the $n+1$th source position with the $n$th source position in the chain, accepting the $n+1$th source position if the acceptance ratio $A(\boldsymbol{\beta}_{n+1},\boldsymbol{\beta}_n)$ has increased from the previous chain. The acceptance ratio is defined as ratio between the test and previous source position posterior:
\begin{equation}\label{eq:acceptance}
    A(\boldsymbol{\beta}_{n+1},\boldsymbol{\beta}_n) = \frac{P(\boldsymbol{\beta}_{n+1})}{P(\boldsymbol{\beta}_{n})}
\end{equation}
where $P(\boldsymbol{\beta})$ is a Gaussian sample posterior:
\begin{equation}\label{eq:qsoposterior}
   \ln \left(P(\boldsymbol{\beta})\right) = -\frac{1}{2}\sum\left[\left(\frac{x'_i - x_i}{\sigma_x}\right)^2 + \left(\frac{y'_i - y_i}{\sigma_y}\right)^2 + \left(\frac{\Delta t'_{iC} - \Delta t_{iC}}{\sigma_{\Delta t}}\right)^2  
   \right] 
\end{equation}
where ($x_i$,$y_i$) and $\Delta t_{iC}$ are the observed image positions and QSO time delays, respectively, and $\sigma_x$, $\sigma_y$, and $\sigma_{\Delta t}$ are the corresponding observational uncertainties. We accept $\boldsymbol{\beta}_{n+1}$ as a better fitting source position if its forward projected images and time delays fit the observations better than $\boldsymbol{\beta}_n$. This corresponds to minimizing equation \ref{eq:qsoposterior}. Equation \ref{eq:qsoposterior} is valid for only the QSO, as it is the only source with measured time delays. For all the other sources, we only consider the image positions for the posterior:
\begin{equation}\label{eq:galposterior}
   \ln \left(P(\boldsymbol{\beta})\right) = -\frac{1}{2}\sum\left[\left(\frac{x'_i - x_i}{\sigma_x}\right)^2 + \left(\frac{y'_i - y_i}{\sigma_y}\right)^2 \right] 
\end{equation}

Lastly, we impose a flat prior on each sampled $\boldsymbol{\beta}$ such that $\widetilde{x_s} \in \left[ min(x_s), max(x_s)\right]$ and $\widetilde{y_s} \in \left[ min(y_s), max(y_s)\right]$, where $(x_s,y_s)$ is the sample of all reconstructed source positions from all performed runs.

For this analysis, we perform our Metropolis-Hastings optimization for a chain of 1000 samples. We treat the optimized source positions from this procedure as a simple perturbation of $\boldsymbol{\beta}$ in the overall mean $\tau(\boldsymbol{\theta})$. The result is a largely unchanged $\tau(\boldsymbol{\theta})$ with reconstructed image positions and time delays that better fit the observations. Doing this process allows us to make finer improvements to the source positions and better predict the central QSO time delay.

\subsection{Model Inputs}\label{txt:input}
As described in Section \ref{txt:grale}, {\tt GRALE} only uses multiple image data as input. As a result of this, it is important that the image data is of high quality. While this is true of the 5 QSO images and time delays, it is less certain for the other 3 sources' image identifications (see Section \ref{txt:data}). As a result of the conflicting classifications in the literature, we adopt 2 sets of image identifications to use as input: (1) All images as shown in Table \ref{tab:data} (henceforth referred to as "all im."), (2) A revised set of images where we exclude A1 and A3 (henceforth referred to as "rev im."). In the latter case ("rev im."), we justify the validity of this choice because of the alternative classification of Group A as a single galaxy source, where we have adopted A2 as the 5 multiple images for this case. We should note that this is merely a modelling assumption, and there is no guarantee that the 5 images of A2 actually originate from the same knot in the source. Furthermore, it has been suggested Group A is in reality made of extended arcs \citep{inada03,forestoribio22}, so our use of just A2 seeks to simplify this problem from a more complex extended source analysis. We note that comparing reconstructions from {\tt GRALE} from different subsets of input images has been performed before \citep{liesenborgs09,ghosh23}.

In addition, because the arrival time of the central QSO image E will be affected by the depth of the gravitational potential of the BCG, we decide to also explicitly include the BCG in another analysis for comparison. In this case, we model the BCG as an Elliptic Sersic Lens \citep{keeton01} with mass density:
\begin{equation}\label{eq:sersic}
    \Sigma_S(\boldsymbol{\xi}) = \Sigma_{cen} \exp \left[ -\left(2n-\frac{1}{3}\right) \left(\frac{\boldsymbol{\xi}}{R_e}\right)^{\frac{1}{n}}\right]
\end{equation}
where $\Sigma_{cen}$ is the central mass density, $n$ is the Sersic index, and $\boldsymbol{\xi}$ is the elliptical coordinate position, defined as $\boldsymbol{\xi} = \left(x,\frac{y}{q}\right)$ with $q$ the axis ratio. Since the BCG is an elliptical galaxy, we adopt $n = 4$ for a de Vaucouleurs model. Using the measured ellipticity $e = 0.30 \pm 0.05$ \citep{oguri10}, we fix $e = 0.35$ for our axis ratio, such that $q = 1 - e = 0.65$. The central density can be solved for:
\begin{equation}\label{eq:centraldensity}
    \Sigma_{cen} = M_{tot}\left( \int_0^{\infty} \exp \left[ -\left(2n-\frac{1}{3}\right) \left(\frac{\boldsymbol{\xi}}{R_e}\right)^{\frac{1}{n}}\right] d\boldsymbol{\xi}\right)^{-1}
\end{equation}
where $M_{tot}$ is the total mass in the Sersic lens. We approximate $M_{tot}$ as the dynamical mass $M_{dyn}$ of the BCG, and find that $\Sigma_{cen} \approx 5.95 \times 10^5$ M$_{\odot}$ pc$^{-2}$ (124.17 g cm$^{-2}$). It is worth noting that the input value for $\Sigma_{cen}$ need not be perfectly representative for the BCG as the genetic algorithm will optimize the weight of the Sersic component. Modelling the BCG as a Sersic with these parameters means that the output models from {\tt GRALE} are now hybrid models instead of purely free-form, as galaxy information has been included. As mentioned in Section \ref{txt:grale}, the inclusion of this Sersic serves as an additional basis function on top of the grid of Plummers that will be optimized alongside the rest of the grid.

In summary, we generate a total of 4 lens models: (1) A free-form reconstruction with no Sersic model for the BCG  and all images included ("no sersic / all im."), (2) A free-form reconstruction with no Sersic model for the BCG and the revised image subset ("no sersic / rev. im."), (3) A hybrid reconstruction with a Sersic model for the BCG and all images included ("w sersic / all im."), (4) A hybrid reconstruction with a Sersic model for the BCG and the revised image subset ("w sersic / rev. im."). We refer to these as our 4 models for this paper.

\section{Results}\label{txt:results}

For each of the 4 lens models in our analysis as described in Section \ref{txt:input}, we perform 40 {\tt GRALE} runs each with 10-15 subdivisions, and a maximum of 2900 Plummers. The choice of 40 runs is a result of limitations from computational resources, and is consistent with previous works using {\tt GRALE} \citep{williams19,sebesta19,gho20,ghosh21,ghosh23}. The best fitting subdivisions in each of the 40 runs are averaged to obtain the lens model for each of the 4 models in our analysis. Using these, we perform the source position optimization (see Section \ref{txt:optim}) to better match image positions of each source and the QSO time delays. These results are analyzed below.

\subsection{Surface Mass Density Distribution}\label{txt:surfdensmaps}

\begin{figure*}
\begin{multicols}{2}
    \includegraphics[trim={5cm 0.37cm 4.7cm 0.37cm},clip,width=0.49\textwidth]{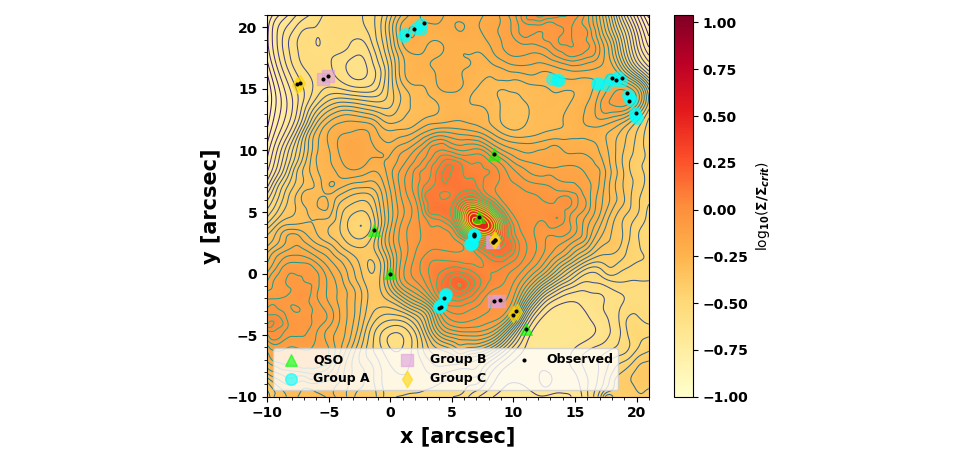}\par 
    \includegraphics[trim={5cm 0.37cm 4.7cm 0.37cm},clip,width=0.49\textwidth]{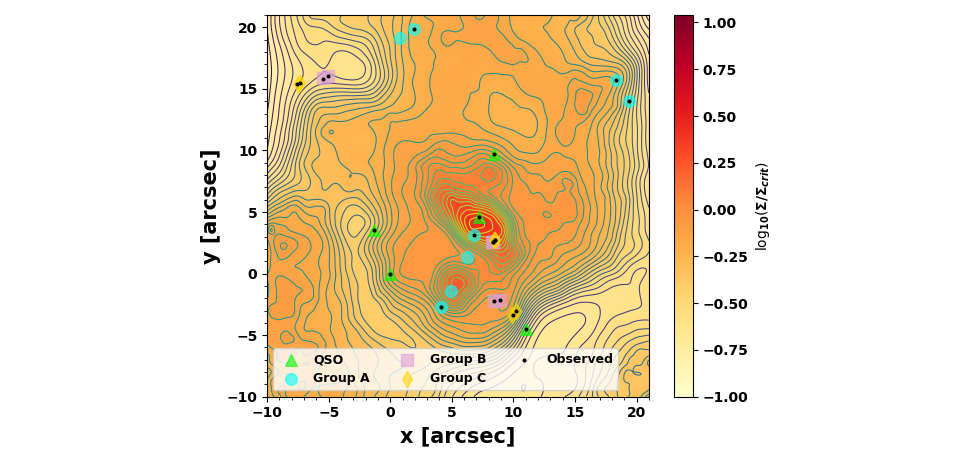}\par 
    \end{multicols}
\begin{multicols}{2}
    \includegraphics[trim={5cm 0.37cm 4.7cm 0.37cm},clip,width=0.49\textwidth]{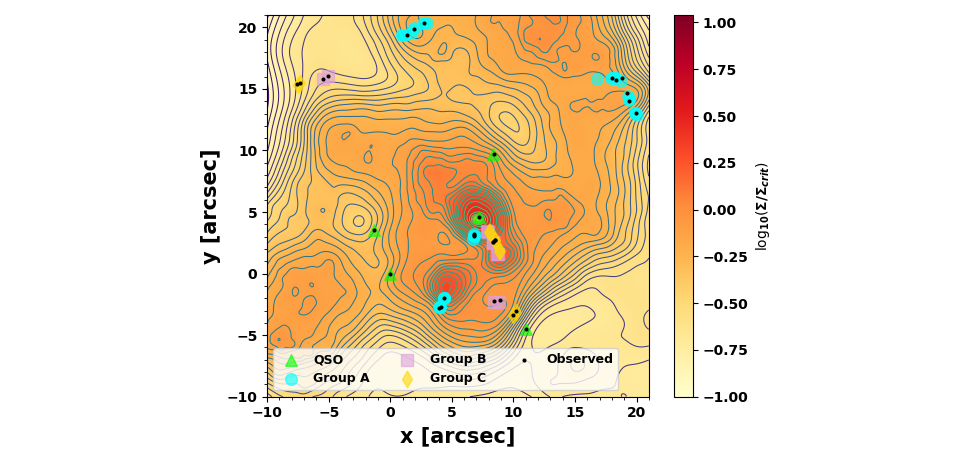}\par
    \includegraphics[trim={5cm 0.37cm 4.7cm 0.37cm},clip,width=0.49\textwidth]{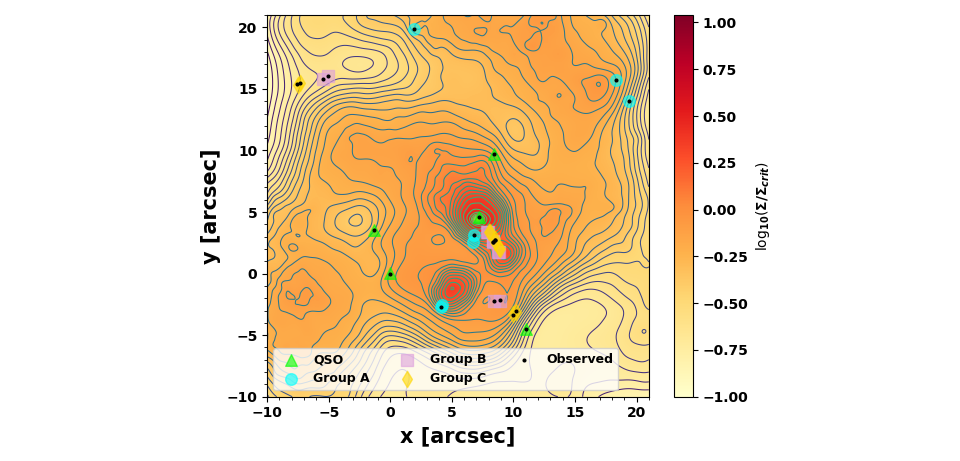}\par
\end{multicols}
\caption{ The {\tt GRALE} reconstructed projected surface mass density distributions for SDSS J1004+4112 for each of our models (see Section \ref{txt:input}): "no sersic / all im." ({\it Top Left}), "no sersic / rev im." ({\it Top Right}), "w sersic / all im." ({\it Bottom Left}), "w sersic / rev im." ({\it Bottom Right}). Images presented are each model's reconstructed multiple images for the 4 sources in this lens. Labels and colours are the same as those in Figure \ref{fig:SDSSJ1004}. Black dots indicate the observed image positions. Contours are spaced logarithmically. For each lens model, 1 arcsec is equivalent to 7.16 kpc. $\Sigma_{\rm crit}$ is measured for the QSO ($z = 1.734$) to be $1.17 \times 10^{11} M_{\odot}$ arcsec$^{-2}$.}
\label{fig:massrecon}
\end{figure*}

\begin{figure*} 
\begin{multicols}{2}
    \includegraphics[width=0.49\textwidth]{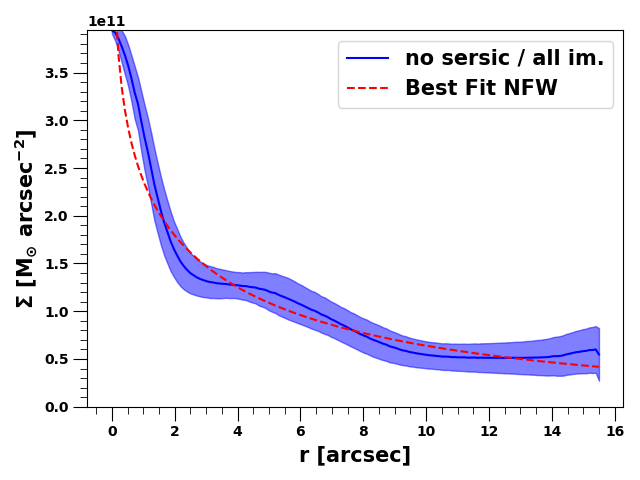}\par 
    \includegraphics[width=0.49\textwidth]{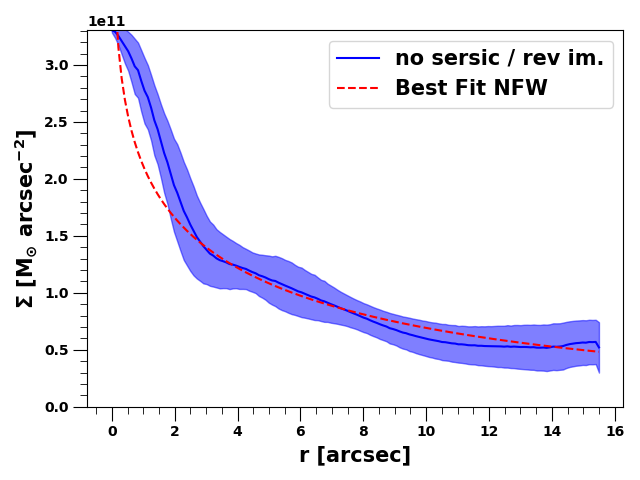}\par  
    \end{multicols}
\begin{multicols}{2}
    \includegraphics[width=0.49\textwidth]{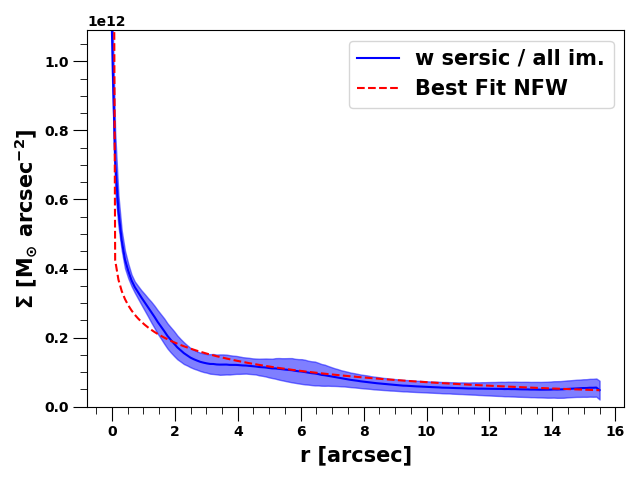}\par 
    \includegraphics[width=0.49\textwidth]{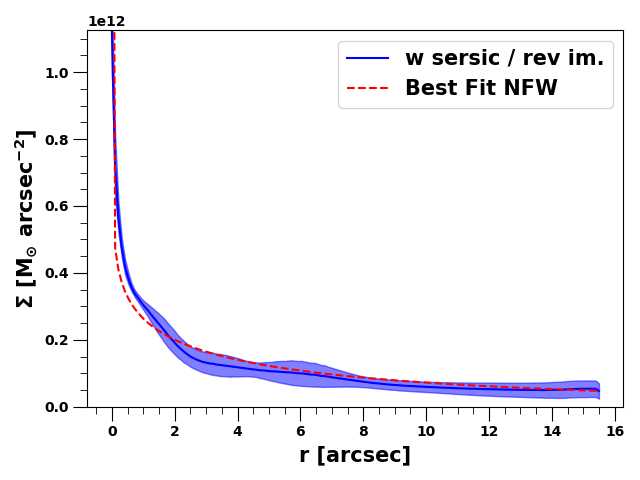}\par  
\end{multicols}
\caption{Circularly Averaged Surface Density Profiles (with respect to the location of the BCG) in blue for each of our models (see Section \ref{txt:input}): "no sersic / all im." ({\it Top Left}), "no sersic / rev im." ({\it Top Right}), "w sersic / all im." ({\it Bottom Left}), "w sersic / rev im." ({\it Bottom Right}). The blue shaded regions are the 68\% confidence level for each reconstructed mass distribution. The dashed red line indicates the best fitting NFW profile (see equation \ref{eq:nfw}).}
\label{fig:circdensprof}
\end{figure*}

\begin{table*}
	\caption{Lens Model Image Reconstructions and NFW Parameters}
	\begin{tabular}{lllllllc} 
		\hline
		Lens Model & $\langle \Delta \boldsymbol{\theta}_{QSO} \rangle$ & $\langle \Delta \boldsymbol{\theta}_{A} \rangle$ & $\langle \Delta \boldsymbol{\theta}_{B} \rangle$ & $\langle \Delta \boldsymbol{\theta}_{C} \rangle$ & $c$ & $R_s$ [arcsec] & $R_{200}$ [kpc] \\
		\hline
		"no sersic / all im." & 0.0119" & 0.1250" & 0.0430" & 0.0465" & 7.06 $\pm$ 0.19 & 21.85 $\pm$ 0.95 & 1105 $\pm$ 58  \\
		"no sersic / rev im." & 0.0109" & 0.0285" & 0.0251" & 0.0298" & 5.36 $\pm$ 0.15 & 32.55 $\pm$ 1.57 & 1247 $\pm$ 61 \\
		"w sersic / all im." & 0.0139" & 0.0932" & 0.0527" & 0.0506" & 6.44 $\pm$ 0.19 & 25.97 $\pm$ 1.24 & 1197 $\pm$ 69 \\
		"w sersic / rev im." & 0.0109" & 0.0491" & 0.0495" & 0.0672" & 7.29 $\pm$ 0.21 & 22.37 $\pm$ 1.06 & 1166 $\pm$ 61 \\
		\hline
	\end{tabular}\\
\medskip{The columns list the lens model (see Section \ref{txt:input}), mean reconstructed image separations from observation (not including extraneous images when applicable) post-source position optimization for the QSO ($\langle \Delta \boldsymbol{\theta}_{QSO} \rangle$), Group A ($\langle \Delta \boldsymbol{\theta}_{A} \rangle$), Group B ($\langle \Delta \boldsymbol{\theta}_{B} \rangle$), and Group C ($\langle \Delta \boldsymbol{\theta}_{C} \rangle$) images, best fitting concentration parameter $c$, NFW scale radius $R_s$, and virial radius $R_{200}$.}
\label{tab:massdensresults}
\end{table*}

Figure \ref{fig:massrecon} shows the surface mass density distributions for all 4 of our lens models. Figure \ref{fig:circdensprof} shows the circularly averaged density profiles for each lens model with each one's best fitting NFW density profile, written as \citep{bartelmann96,nfw97,wright00}:
\begin{equation}\label{eq:nfw}
    \Sigma_{NFW}(x) = \left\{ \begin{array}{lrc} \frac{2 R_s \delta_c \rho_c}{x^2 - 1}\left( 1 - \frac{2}{\sqrt{1 - x^2}}\tanh^{-1}\left(\sqrt{\frac{1-x}{1+x}}\right)\right) & \mbox{for} & x<1 \\ \frac{2 R_s \delta_c \rho_c}{3} & \mbox{for} & x = 1 \\
    \frac{2 R_s \delta_c \rho_c}{x^2 - 1}\left( 1 - \frac{2}{\sqrt{x^2 - 1}}\tan^{-1}\left(\sqrt{\frac{x-1}{1+x}}\right)\right) & \mbox{for} & x>1
\end{array}\right.
\end{equation}
where $c$ is the concentration parameter, $R_s$ is the scale radius, $x$ is the scaled distance from the NFW center $(x = r/R_s)$, $\delta_c$ is the characteristic overdensity, and $\rho_c$ is the critical density of the universe. In fitting the circularly average density profiles to a NFW, we can estimate the virial radius of the cluster $R_{200} = R_s c$ using the best fit $c$ and $R_s$. We note that while interesting to see if a lens model's mass density profile follows an NFW profile, we recognize that such a fit is not very physically accurate as the mass density distribution is not circularly symmetric. Nonetheless, the result serves as a useful comparison with Chandra X-ray observations ($R_s = 39^{+12}_{-9}$ arcsec, $c = 6.1^{+1.5}_{-1.2}$) \citep{ota06}, and  parametric models \citep{oguri10,forestoribio22}. Table \ref{tab:massdensresults} summarizes each model's image reconstruction properties and best fitting NFW parameters.

\begin{figure}
    \centering
    \includegraphics[trim={5cm 0.37cm 4.5cm 0.37cm},clip,width=0.49\textwidth]{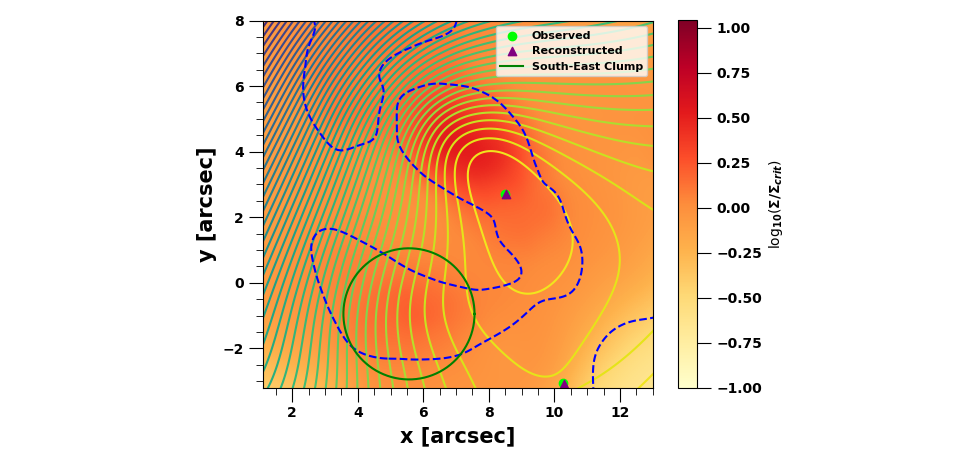}
    \includegraphics[trim={5cm 0.37cm 4.5cm 0.37cm},clip,width=0.49\textwidth]{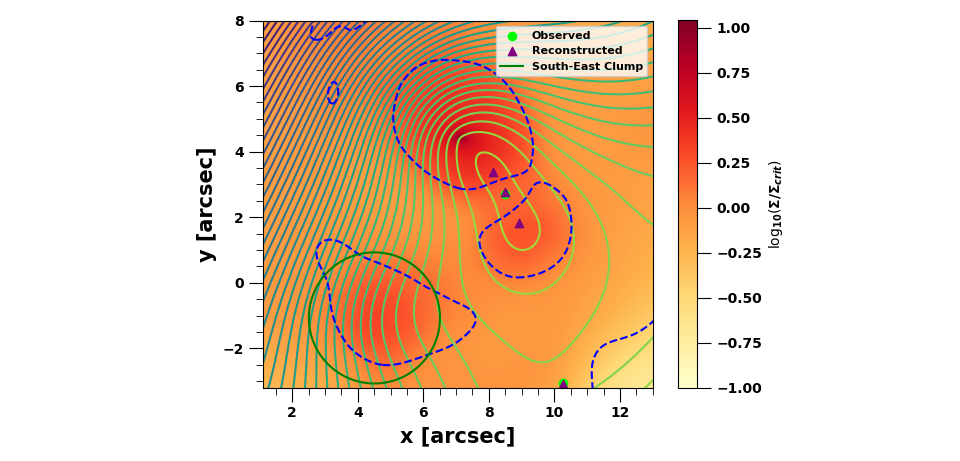}
\caption{Zoomed in view of the central density distribution of "no sersic / all im." ({\it Top}) and "w sersic / all im." ({\it Bottom}). Lime green dots and purple triangles represent the observed and reconstructed image positions for source C1. Contours represent the time delay surface $\tau(\boldsymbol{\theta})$, where stationary points correspond with reconstructed image positions. Dashed blue lines are the critical curves for Group C ($z_s = 3.28$). The green circle indicates the south-east clump, with a radius of $2"$ centered on the mass peak of the clump. It is noted that the clump is coincident on a critical curve, whose presence explains the extraneous images formed in AX.1 and AX.5 (X $\in [1,3]$). The main feature of the figure is the steeper mass peak (representing the BCG) present in the "w sersic / all im." model that splits the original ({\it Top}) critical curve into 3 sections, thus forming 2 extraneous central images. While only shown here for source C1, this effect is present in all the hybrid model reconstructions for the central images of Groups B and C. 1 arcsec is equivalent to 7.16 kpc, and $\Sigma_{\rm crit}$ is measured for the QSO ($z = 1.734$) to be $1.17 \times 10^{11} M_{\odot}$ arcsec$^{-2}$.}

\label{fig:centraldist}
\end{figure}

\begin{figure*}
\begin{multicols}{2}
    \includegraphics[width=0.49\textwidth]{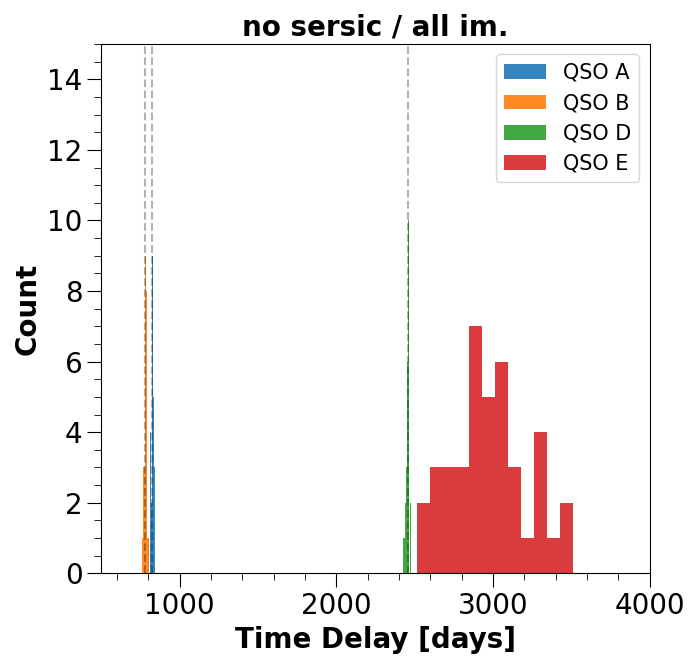}\par 
    \includegraphics[width=0.49\textwidth]{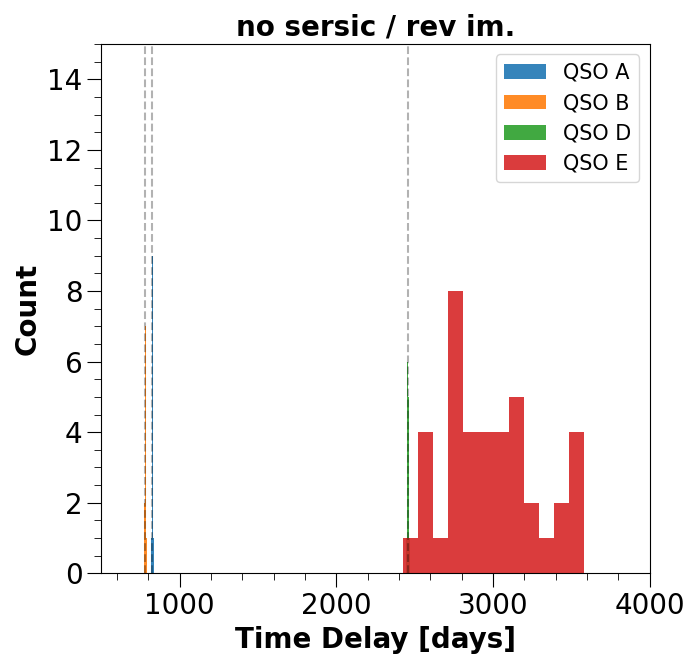}\par 
    \end{multicols}
\begin{multicols}{2}
    \includegraphics[width=0.49\textwidth]{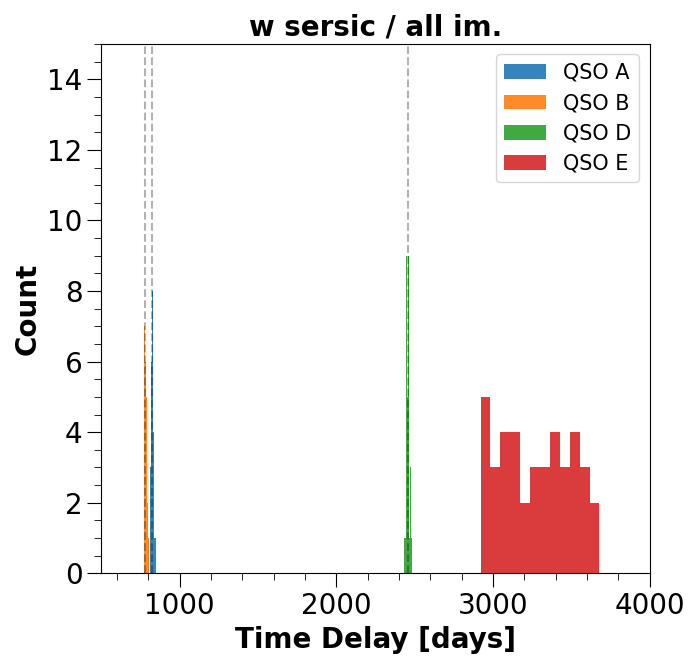}\par
    \includegraphics[width=0.49\textwidth]{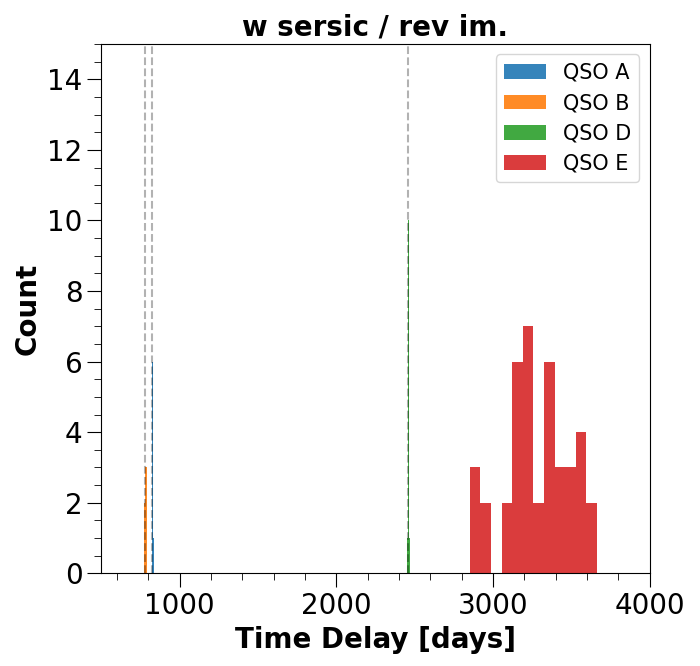}\par
\end{multicols}
\caption{Histograms of QSO time delay measurements of the 40 {\tt GRALE} runs for each of our lens models (see Section \ref{txt:input}): "no sersic / all im." ({\it Top Left}), "no sersic / rev im." ({\it Top Right}), "w sersic / all im." ({\it Bottom Left}), "w sersic / rev im." ({\it Bottom Right}). All time delays are measured with respect to QSO image C. Dashed grey lines indicate the observed time delay for QSO image B, A, and D, respectively, from left to right. Though not shown in the figure, final time delay measurements (see Table \ref{tab:tdelayresults}) for each lens model are calculated by averaging the time delay surfaces of the 40 {\tt GRALE} runs (shown here for each image), then performing our source position optimization (see Section \ref{txt:optim}).}
\label{fig:tdelays}
\end{figure*}

\begin{table*}
	\caption{Time Delay Results}
	\begin{tabular}{llllllc} 
		\hline
		QSO Image & Observed $\Delta t$ & "no sersic / all im." $\Delta t$ & "no sersic / rev im." $\Delta t$ & "w sersic / all im." $\Delta t$ & "w sersic / rev im." $\Delta t$ &  \\
		\hline
		A & 825.99 $\pm$ 2.10 & 824.96 $\pm$ 9.05 & 827.33 $\pm$ 3.81 & 825.09 $\pm$ 9.48 & 826.90 $\pm$ 3.95 &  \\
		  B & 781.92 $\pm$ 2.20 & 781.31 $\pm$ 8.68 & 783.26 $\pm$ 3.69 & 781.55 $\pm$ 9.16 & 782.36 $\pm$ 3.67 & \\
		D & 2456.99 $\pm$ 5.55 & 2458.19 $\pm$ 9.61 & 2458.17 $\pm$ 5.07 & 2459.03 $\pm$ 10.36 & 2458.39 $\pm$ 5.23 & \\
		E & ---- & 2974.04 $\pm$ 242.18 & 2990.79 $\pm$ 294.02 & 3275.73 $\pm$ 223.79 & 3282.71 $\pm$ 203.68 & \\
		\hline
	\end{tabular}\\
\medskip{ QSO time delay measurements in days for each lens model. All time delays $\Delta t$ are with respect to QSO image C, the first arriving image, and calculated after averaging over 40 independent {\tt GRALE} runs and performing our source position optimization (see Section \ref{txt:optim}). }
\label{tab:tdelayresults}
\end{table*}

\subsubsection{Free-Form {\tt GRALE} Models}\label{txt:freeform}
The top row of Figure \ref{fig:massrecon} shows our 2 free-form mass density maps for "no sersic / all im." (left) and "no sersic / rev im." (right). Both models generally reconstruct mass features consistent with those found in other reconstructions for this lens, particularly orientation of the central distribution about the BCG and a mass concentration in the south-east, at $\sim(5'',-1'')$, near A3.1 \citep[e.g.][]{liesenborgs09,mohammed15,forestoribio22,liu23}, though the shape of this clump varies between different reconstructions, and often manifests as an extension of mass from the center of the cluster. For example, in \cite{forestoribio22} it is represented by the multipole potential term which has a mass extension along this polar direction with respect to the BCG. In our reconstruction, this south-east clump has a radius of $\sim 2"$ containing $\sim 3.4 \times 10^{11} M_{\odot}$ (with background density subtracted) and is not closely affiliated with any observed cluster galaxy members \citep{oguri10}. Figure \ref{fig:centraldist} shows a zoomed in view of the central region for both free-form models, with the critical curves (dashed blue lines) and time delay surface contours (green solid lines) plotted over the mass density maps. As shown in Figure \ref{fig:centraldist}, the south-east clump also significantly influences the critical curve. Also common to both models is the total mass within 60 kpc (roughly the region bounded by the QSO images), found to be $\sim 2.5 \times 10^{13} M_{\odot}$,  which is consistent with past results \citep{williams04}.

The "no sersic / all im." model has a central peak displaced about 0.2" (1.4 kpc) from the BCG. This very small displacement from the BCG indicates that {\tt GRALE} can localize galaxy centers quite accurately if nearby images are present. The central mass density peak corresponds to $\sim 4 \times 10^{11} M_{\odot}$ arcsec$^{-2}$, which is about 3 times lower than when the BCG is explicitly included (see Section~\ref{txt:hybrid}).
This model reconstructs the QSO image positions to within the HST accuracy, with mean image separation from the observed images of $\langle \Delta \boldsymbol{\theta}_{QSO} \rangle = 0.0119"$. The circularly averaged density profile is shown in the top left of Figure \ref{fig:circdensprof}. The best fit NFW has $R_s = 21.85 \pm 0.95$ arcsec ($156.4 \pm 7.0$ kpc) and $c = 7.06 \pm 0.19$. This yields a virial radius of 1105 $\pm$ 58 kpc. While this $c$ is consistent with that found by \cite{ota06}, this model's $R_s$ is in disagreement. Parametric models find $c$ and $R_s$ measurements disagree with \cite{ota06}. These models exhibit degeneracies of these parameters within their lens models, and are in fact consistent with \cite{ota06} when considering mass within $\sim$100 kpc to break the degeneracy \citep{oguri10,forestoribio22}. Similarly, for this model, we find $\sim 4.6 \times 10^{13} M_{\odot}$ within 100 kpc, consistent with the mass of $5.0^{+1.8}_{-1.1} \times 10^{13} M_{\odot}$ estimated by \cite{ota06}.

For Groups A, B, and C, images are reconstructed to better than $0.13''$, i.e., within the observed positional uncertainty of $0.4"$, as estimated by \cite{oguri10}. In general, this model reconstructs image positions well despite also generating extraneous images. As seen in the top left of Figure \ref{fig:massrecon}, a number of extraneous images are produced for Group A. Specifically, each reconstructed image of A1, A2, and A3 lies very near to the critical curve, thus producing extraneous images close to the actual positions of each image \citep[also seen in][]{forestoribio22,liu23}. In the case of AX.2, AX.3, and AX.4 (X $\in [1,3]$), these extraneous images form in a very similar way to those found in \cite{forestoribio22}, with the critical curves "unfolding" the individual reconstructed image into three. These extraneous images form with similar time delays and magnifications as their observed reconstructed image, due to the local time delay surface remaining relatively smooth. In the case of AX.1 and AX.5 (X $\in [1,3]$) the extraneous images that form can be explained in Figure \ref{fig:centraldist} by the south-east clump forming a critical curve that splits AX.1 and creates a steeper central mass profile whose critical curve splits AX.5. This is something present in all 4 of our models. Harking back to the image classification controversy in the literature (see Section \ref{txt:data}) and the large uncertainty in Group A's image positions, the number of extraneous images is strongly affected by the chosen input images, and we note that the number of extraneous images is significantly reduced when ignoring A1 and A3. In fact, the aforementioned "unfolding" largely disappears when we use the revised image subset. Interestingly, the extraneous images of this model roughly reproduce the observed image positions of AX.4, which form a kink relative to AX.3 (X $\in [1,3]$), in the upper right of Figure~\ref{fig:SDSSJ1004}. We hypothesize that all of this is indicative of the necessity to treat Group A as an extended source rather than a series of point sources because the image knots that make up Group A might have been misidentified. For now, though, we can ignore these extraneous images as they form very similar time delays and magnifications as their individual respective reconstructed images.

The "no sersic / rev im." model has a central peak displaced about 0.3" (1.9 kpc) from the BCG. The central mass density peak corresponds to $\sim 3 \times 10^{11} M_{\odot}$ arcsec$^{-2}$. Overall, this lens model has a flatter mass density distribution than the "no sersic / all im." model. The best fit NFW profile (top right of Figure \ref{fig:circdensprof}) has $c = 5.36 \pm 0.15$ and $R_s = 32.55 \pm 1.57$ arcsec ($232.5 \pm 9.8$ kpc), both of which are in good agreement with those from NFW fits to the density profile derived from Chandra X-ray observations. This yields a virial radius of 1247 $\pm$ 61 kpc. Similar to the "no sersic / all im." model, this model reconstructs the QSO image positions very well, with $\langle \Delta \boldsymbol{\theta}_{QSO} \rangle = 0.0109"$. 

Unique to this model is that all images in Group A, B, and C are reconstructed to within HST accuracy. Furthermore, this model found the fewest extraneous images, all for Group A. Of the few extraneous images, all were formed very close to the critical curve as before. Notably, A2.3, and A2.4 do not form extraneous images. From this result, and the fact that "no sersic / rev im." model had the best reconstructed image positions of all our models, we conclude that the use of single point images for Group A (in our case just A2) reduces the prevalence of extraneous images in models for SDSS J1004+4112. An extended source analysis for Group A would likely be the most accurate, because the current identification of knots as belonging to the same source(s) may be incorrect. Nonetheless, both the "no sersic / rev im." model and "no sersic / all im." model are able to reconstruct the same main features of the cluster.

\subsubsection{Hybrid {\tt GRALE} Models}\label{txt:hybrid}
The bottom row of Figure \ref{fig:massrecon} shows our 2 hybrid mass density distributions for "w sersic / all im." (left) and "w sersic / rev im." (right). As with the free-form models, both hybrid models also have $\sim 2.5 \times 10^{13} M_{\odot}$ within 60 kpc, indicating that all four of our models reproduce the central mass distribution consistently with one another. The hybrid models also reconstruct the same mass features as the free-form models. The south-east clump forms in roughly the same place, and contains $\sim 6.0 \times 10^{11} M_{\odot}$ in both hybrid models. The main difference between the hybrid models and the free-form models is that the hybrid models' mass distributions are much more strongly peaked at the location of the BCG, which is a consequence of the inclusion of a Sersic profile to model the BCG. Naturally, both hybrid models have a central peak at the BCG location. Since the mass distribution is much steeper in the central region, a smaller south-west mass clump located $\sim$(9.0",1.5") is also visible, containing $\sim 3.5 \times 10^{11} M_{\odot}$ in both hybrid models. This south-west mass clump appears to also be present in the free-form models, but is less visible in the density profile due to the free-form mass distribution being much flatter. Notably, unlike the south-east clump, the south-west clump appears to be consistent with a cluster member galaxy located at ($9.36",2.41"$) \citep{oguri10}.

The "w sersic / all im." model has a central mass density peak of $\sim 1.3 \times 10^{12} M_{\odot}$ arcsec$^{-2}$. Just as the free-form models, this model reconstructs the QSO image positions well, with $\langle \Delta \boldsymbol{\theta}_{QSO} \rangle = 0.0139"$. Likewise, the images for Group A, B, and C are also reconstructed to within positional uncertainty, as estimated by \cite{oguri10}. The bottom left of Figure \ref{fig:circdensprof} shows the best fitting NFW profile with $c$ and $R_s$ found to be 6.44 $\pm$ 0.19 and 25.97 $\pm$ 1.24 arcsec ($185.5 \pm 9.1$), respectively. This yields a virial radius of 1197 $\pm$ 69 kpc. $R_s$ is underestimated in this model compared to \cite{ota06}. The mass within 100 kpc is measured to be $\sim 4.6 \times 10^{13} M_{\odot}$, consistent with Chandra X-ray observations. 

Like its free-form counterpart ("no sersic / all im."), the "w sersic / all im." model finds a number of extraneous images. These are primarily formed for Group A in the same way as before. Unique to both hybrid models is the formation of extraneous images very close to the central images of Group B and C (BX.3 and CX.3). This is caused by the critical curve forming very close to those images in these models, likely the result of the much steeper mass distribution splitting the critical curve at that location, as demonstrated in Figure \ref{fig:centraldist}. Indeed the time delay surface near BX.3 and CX.3 is only slightly perturbed to form the central image very near to the observed BX.3 and CX.3 position, with 2 small $\sim 1.04 \Delta \tau_{32}$ (where $\tau_{32}$ is the time delay between BX.3/CX.3 and BX.2/CX.2) peaks forming demagnified extraneous images. Therefore, it is implied that the extraneous images of Group B and C can be ignored as they form very close to one another and form nearly the same time delays, similar to the extraneous Group A images discussed in Section \ref{txt:freeform}. We also hypothesize that these extraneous images could be removed by including the cluster member galaxy at ($9.36",2.41"$) as a prior in a future {\tt GRALE} run, as mass currently concentrated at the BCG would be spread out, thus smoothing out the mass distribution and reforming the split critical curve. The result also suggests that Group B and C may also be more accurately described by an extended source.

The "w sersic / rev im." model also has a central mass density peak of $\sim 1.3 \times 10^{12} M_{\odot}$ arcsec$^{-2}$ and mass within 100 kpc measured to be $\sim 4.6 \times 10^{13} M_{\odot}$, consistent with Chandra X-ray observations. This model also reconstructs the QSO image positions well, with $\langle \Delta \boldsymbol{\theta}_{QSO} \rangle = 0.0109"$. The best fit NFW profile (bottom right of Figure \ref{fig:circdensprof}) finds $c$ and $R_s$ to be 7.29 $\pm$ 0.21 and 22.37 $\pm$ 1.06 arcsec ($159.8 \pm 7.1$ kpc), respectively. This yields a virial radius of 1166 $\pm$ 61 kpc.

Similar to its free-form counterpart ("no sersic / rev im."), the "w sersic / rev im." model reproduces the images of Group A, B, and C better and has fewer extraneous images than in the models that use all images. This model is unique in that it is the only model to not produce extraneous images for A2.2, while A2.1 and A2.5 still lie very close to the critical curve and produce extraneous images. Similarly, BX.3 and CX.3 also form 2 demagnified extraneous images just as in the "w sersic / all im." model. Regardless, both hybrid models properly reconstruct all the main observational features of cluster.

\subsubsection{The South-East Mass Clump}\label{txt:clump}
The south-east mass clump's consistent presence in all 4 of our lens models, disaffiliation with any cluster galaxy members, and apparent presence in previous lens reconstructions makes it an intriguing candidate for possible dark matter substructure. This requires strong evidence that the clump is indeed necessary in order to reproduce the observable features of the lens. As it stands, there is not enough evidence to definitively constrain the south-east clump's mass distribution, largely stemming from the lack of observational constraints available in SDSS J1004+4112. The clump's location between AX.1 and AX.5 subjects it to the monopole degeneracy allowing its mass to be redistributed within that region without changing the image positions or time delays, thus smoothing out the mass distribution. This is a fundamental degeneracy in gravitational lensing and is generally difficult to break without a large number of sources and high image density \citep{liesenborgs08}. Since SDSS J1004+4112 lacks both of these prerequisites, it can be argued that the south-east clump can simply be redistributed in the region between AX.1 and AX.5. However, its location very near to AX.1 limits the scale at which one can apply the monopole degeneracy, making it likely that the clump may not be fully redistributed. The south-east clump ultimately needs further study in order to verify its existence.

As mentioned previously, excess mass in the same vicinity of the south-east clump does seem to be present in previous reconstructions. In particular, the multipole perturbation in the model presented in \cite{forestoribio22} finds a density peak at $\sim 230$ degrees from the west axis, coincident with the rough location of the south-east clump in our model. Likewise we perform a separate lens reconstruction, not shown in a figure in this article, which includes the nearest cluster member galaxy as an additional basis function in the "w sersic / all im." model. We find that the south-east clump persists in this case with roughly the same mass and size, implying that if it exists it is not affiliated with any galaxy. For these reasons there does seem to be some mass feature in this region of the cluster, although whether or not the south-east clump is the solution cannot be confirmed. 

Interestingly, there is actually a galaxy visible in the location of the south-east clump (visible just above AX.1 in Figure \ref{fig:SDSSJ1004}). This galaxy is notably bluer than all associated cluster members, thus likely rendering it not a cluster member galaxy but rather a foreground galaxy.  Depending on the precise redshift of this structure, this could contribute to the lensing signal as a line of sight structure, suggesting that what was called the "dark" south-east clump perhaps also has a visible component to it. If the galaxy is foreground and the south-east clump is associated with it, the mass of the clump will need to be larger because of the larger $\Sigma_{\rm crit}$ for most smaller $z_d$'s. To our knowledge, unfortunately no redshift is available for said galaxy.

If we assume the south-east clump is in fact real, then it would imply that the cluster may not be fully relaxed. Additionally, it can be used in the future to provide constraints on the nature of dark matter, such as its interaction cross-section per unit mass, $\sigma/m$. This has been done before with dark matter substructures in cluster lenses using sophisticated cosmological and N-body simulations \citep{peter13,harvey19,xu23} and strong lensing analysis of existing lenses \citep{miraldaescude02,markevitch04,bradac08,andrade22}. While a more complex substructure analysis for SDSS J1004+4112 would be the subject of a future paper, we can perform a back-of-envelope constraint for $\sigma/m$ using the south-east clump, assuming dark matter is self-interacting, by following the exercise done by \cite{ghosh23}. 

First, if we assume that the clump is gravitationally bound and virialized, we can calculate the typical particle velocity $v$ using the virial theorem (with $M$ the clump mass and $r$ the radial size of the clump, assumed to be $2"$ (14.3 kpc) for both models): $v \sim GM/r$. For free-form and hybrid models we calculate $v$ as $\sim 320$ km s$^{-1}$ and $\sim 420$ km s$^{-1}$, respectively. We note that the clump mass is rather large (on the order of $10^{11} M_{\odot}$), resulting in a large density $\sim 10^5 \rho_{c}$. This is not surprising as the clump is located in the central region of the cluster. The relatively large $v$ that we have calculated makes sense since a large velocity dispersion would be required to hold the clump up against self-gravity, given its small size. As a self-consistency check of this gravitationally bound assumption, and of the necessarily large $v$, we calculate the time to cross the clump's diameter at this velocity to be $<9 \times 10^7$ years. Because this time scale is short on cosmological scales, we conclude that unless we happen to be observing the lens at a very special time, the clump, if real, is most likely gravitationally bound.

The preceding paragraph makes an implicit assumption that the dark matter particles are non-interacting. The apparent longevity of the clump allows us to place an upper limit on the self-interaction cross-section of the dark matter. Using the calculated $v$, we can calculate the particle dispersal time $\tau$, which is the time for the substructure to completely disperse due to scattering induced by self-interactions \citep{miraldaescude02,peter13}, as $\tau = [\rho (\sigma / m) v]^{-1}$. Here we assume the clump is spherical in 3D with mean density $\rho$. If we assume $\sigma/m = 4.0$ cm$^2$ g$^{-1}$, the dispersal time is $\sim 1.3 \times 10^8$ years and $\sim 5.8 \times 10^7$ years for free-form and hybrid models, respectively. Assuming the cluster formed at $z = 1$, this represents $\sim 8\%$ and $\sim 4\%$ of the time between formation and observation at $z_d = 0.68$ for free-form and hybrid models, respectively. Since $\tau$ is so short, this suggests that $\sigma/m = 4.0$ cm$^2$ g$^{-1}$ is a rough upper limit as a smaller cross section would imply a longer lifespan of the clump. This constraint is consistent with those found from the Bullet Cluster \citep{markevitch04}, and MACS J0025 \citep{bradac08} using the same method. Using $\sigma/m = 1.0$ cm$^2$ g$^{-1}$, a more restrictive cross section constraint, we find $\tau$ $\sim 5.1 \times 10^8$ years and $\sim 2.5 \times 10^8$ years for free-form and hybrid models, respectively, representing $\sim 32\%$ and $\sim 16\%$ of the time between formation and observation, respectively. While there is a longer dispersal time in this scenario, $\tau$ is still relatively short compared to the assumed lifetime of the cluster, so we can tentatively assign an upper limit of $\sigma/m = 1.0$ cm$^2$ g$^{-1}$ for the self-interacting cross section of dark matter using just the south-east clump, assuming it is a real mass structure. A future, more rigorous analysis is required to get a more stringent constraint on $\sigma/m$. For now, this back-of-envelope calculation shows the utility of using dark matter substructures in clusters to constrain dark matter self-interaction properties.

Recently, wave dark matter (also known as fuzzy dark matter) has had success in reproducing observed flux ratios in strong lenses \citep{amruth23,powell23}, leading to burgeoning interest as a candidate for dark matter. For both self-interacting and wave dark matter, a more detailed study would produce models using constraints given from the macrolens model, such as this work. As an order of magnitude estimate, we can calculate the de Broglie wavelength $\lambda$ assuming wave dark matter using the scaling relation from \cite{schive14}: $\lambda \propto m_{\psi}^{-1} M^{-1/3}$ where $M$ is the mass of the dark matter halo and $m_{\psi}$ is the mass of the wave dark matter particle. In wave dark matter, $\lambda$ sets the length scale at which the density distribution of dark matter will fluctuate, producing what are commonly referenced to as granules. These granules' effect on gravitational lensing has been well studied both with simulations and modelling from observations \citep{dalal21,laroche22,amruth23,powell23,diego23}. With the south-east clump as a test case, and assuming the typically quoted $m_{\psi} \sim 10^{-22}$ eV, we estimate $\lambda$ as $\sim 0.21$ kpc and $\sim 0.18$ kpc for free-form and hybrid models, respectively. This scale is too small to be resolved by HST for SDSS J1004+4112. It is nonetheless consistent with the expected range necessary to produce a solitonic core \citep{amruth23}, which is apparently a feature of wave dark matter \citep{schive14}. Like our calculation for self-interacting dark matter, we note the utility in using substructure in clusters to constrain models of dark matter, and encourage future detailed studies utilizing substructure constraints to formalize properties of dark matter.

\subsection{Time Delays}

Figure \ref{fig:tdelays} presents histograms of the QSO time delay measurements for the 40 {\tt GRALE} runs for each of our 4 lens models. At this step, we can see that the measured time delays for the 3 images with observed time delays (QSO images A, B, and D) are reconstructed with very small variance. The models using the revised set of images (ignoring A1 and A3) calculate tighter constraints on the time delays for QSO images A, B, and D.

As described in Section \ref{txt:results}, we average over the 40 {\tt GRALE} runs' time delay surfaces, then perform the source position optimization to obtain the final lens models. Source position optimization is treated as a simple perturbation of $\boldsymbol{\beta}$, and this is reflected by the fact that the optimized time delays differ from the initial mean time delay of the 40 {\tt GRALE} runs by $\sim$1-2 days. The optimized time delays for each of the 4 lens models are presented in Table \ref{tab:tdelayresults}.

The time delays for QSO image A, B, and D are all successfully reconstructed to within the observed uncertainty \citep{fohlmeister08,munoz22} in each lens model. The central image, QSO image E, does not have an observed time delay $\Delta t_{EC}$ as of this writing. Using our models, we are able to make 4 predictions for $\Delta t_{EC}$. Interestingly, the free-form and hybrid {\tt GRALE} models predict different $\Delta t_{EC}$ and are broadly in tension with one another. The free-form models predict $\Delta t_{EC}$ of 2974.04 $\pm$ 242.18 days and 2990.79 $\pm$ 294.02 days for "no sersic / all im." and "no sersic / rev im.", respectively. The hybrid models predict longer $\Delta t_{EC}$ of 3275.73 $\pm$ 223.79 days and 3282.71 $\pm$ 203.68 days for "w sersic / all im." and "w sersic / rev im.", respectively. Only the predicted "no sersic / rev im." $\Delta t_{EC}$ is consistent with the hybrid model predictions. In general, the longer time delay predictions for the hybrid models makes sense as the inclusion of the Sersic profile at the location of the BCG means light from the QSO source will travel through a deeper potential.

For the recently measured time delay of QSO image D $\Delta t_{DC}$ (see Table \ref{tab:tdelayresults}) \citep{munoz22}, only \cite{mohammed15} had a successful prediction, albeit they had a very broad credible interval between $\sim$1500 and 2700 days. Both \cite{liesenborgs09} ($\Delta t_{DC} \approx 2126$ days) and \cite{oguri10} ($\Delta t_{DC} \approx 2044$ days) underestimated the observed $\Delta t_{DC}$, and likewise predict short $\Delta t_{EC}$ of $\sim 2726$ days and $\sim 2500$ days, respectively. These predictions are mostly inconsistent with our predictions, with only the "no sersic / rev im." $\Delta t_{EC}$ measurement consistent with the prediction from \cite{liesenborgs09}. Our results are consistent with the broad prediction of $\sim 1900 - 3200$ days from \cite{mohammed15}. Our free-form model predictions are consistent with the prediction of 2853.90 days from \cite{forestoribio22}, who used parametric models including the observed $\Delta t_{DC}$.

\section{Discussion and Conclusions}\label{txt:conclusion}

We have presented 4 reconstructions of SDSS J1004+4112 using the free-form genetic algorithm based method {\tt GRALE}. Two of these models are completely free-form with no inclusion of cluster member galaxies, while the other 2 are hybrid models including a Sersic profile to model the BCG. In addition, we have studied the effect of revising the input image data in both mentioned cases to exclude A1 and A3, motivated by a discrepancy in the literature regarding the classification of these images. Our main improvement over previous {\tt GRALE} reconstructions for this lens is the inclusion of the recently measured $\Delta t_{DC}$, which provides stronger constraints on the mass distribution and central image time delay $\Delta t_{EC}$. Our main results are as follows:

\begin{itemize}
    \item All 4 mass distributions (Figure \ref{fig:massrecon}) are broadly consistent with previous studies, both using parametric \citep{oguri10,forestoribio22}, and free-form \citep{williams04,liesenborgs09,mohammed15} methods. The best fit NFW profiles for all 4 mass distributions find concentration parameters $c$ in agreement with those derived from Chandra X-ray observations \citep{ota06}. Even though all but 1 of our models find underestimated best fit NFW scale radii $R_s$, all of our models are consistent with the estimated mass within 100 kpc from \cite{ota06}. Lastly, all input images for each lens model are reconstructed to within the estimated positional uncertainty as used in \cite{oguri10} (see Table \ref{tab:massdensresults}).
    \item To our knowledge, our source position optimization (see Section \ref{txt:optim}) has not been used in previous lens modelling algorithms. Its inclusion in our lens models allows us to better match the observed image positions and time delays using the final {\tt GRALE} mass distributions.
    \item We report 2 main potential substructures: a larger $\sim 2"$ south-east clump at $\sim (5",-1")$ and a smaller south-west clump at $\sim (9.0",1.5")$. The south-west clump is consistent with a cluster member galaxy identified by \cite{oguri10}. The south-east clump is not associated with a cluster member galaxy. This south-east clump is clearly visible in all 4 of our models (containing $\sim 3.4 \times 10^{11} M_{\odot}$ and $\sim 6.0 \times 10^{11} M_{\odot}$ for free-form and hybrid models, respectively) and appears to be present in some form in previous lens reconstructions \citep{liesenborgs09,mohammed15,forestoribio22}. More data and lens models are needed to constrain its mass distribution.
    \item Using a simplified back-of-envelope exercise, we tentatively calculate the upper limit of the dark matter self-interaction cross section to be $\sigma/m = 1.0$ cm$^2$ g$^{-1}$ with the south-east clump, assuming it is real, consistent with constraints from other cluster lenses \citep{markevitch04,bradac08}. A more rigorous analysis of this substructure would be required to confirm if $\sigma/m$ is consistent with more restrictive constraints from recent analyses \citep{harvey19,andrade22}. Instead, if we assume wave dark matter, we calculate its de Broglie wavelength as $\lambda \sim 0.2$ kpc, consistent with the range capable of producing solitonic cores \citep{amruth23}, but also requiring more rigorous analysis.
    \item In comparing our free-form and hybrid models, we find that the hybrid models produce much steeper mass distributions in the central regions. This is largely a result of the inclusion of a Sersic profile at the BCG location. This has the effect of the hybrid models predicting a longer central image time delay, $\Delta t_{EC} \sim 3280$ days, than the free-form prediction of $\Delta t_{EC} \sim 2980$ days. Both of our models predict $\Delta t_{EC}$ consistent with those found by \cite{mohammed15}, which was the only successful prediction of $\Delta t_{DC}$. A future measurement of $\Delta t_{EC}$ will help determine the steepness of the central density profile for this lens.
\end{itemize}

As mentioned, a future measurement of $\Delta t_{EC}$ will help constrain the central density profile as well as provide tighter constraints on all the measured parameters in our study. \cite{forestoribio22} states that this time delay will be too difficult to measure since the central E image is too faint. It should be noted that this presumption is based on visible and IR observations of SDSS J1004+4112 \citep{inada08,munoz22}. Recently \cite{perera23} presented an observing strategy for the central images of quads using UV observations with HST. The QSO source in SDSS J1004+4112 fits the criteria for their strategy, with the source (a QSO) SED peaking in UV wavelengths and the BCG, being an elliptical galaxy, peaking in visible wavelengths. Since the central image is superimposed by the BCG \citep{inada05}, a UV measurement should maximize the contrast between the two, making the central image easier to detect and brighter than in visible wavelengths. In principle, a light curve in UV wavelengths including the central image can be used to potentially measure $\Delta t_{EC}$, with the advantage being that QSO E is more visible in this regime. As a result, we suggest future studies look for the central image time delay with UV observations (or existing archival observations in UV), following the strategy from \cite{perera23}.

Additionally, in order to confirm the reality of the south-east clump, more observational constraints are required. Ideally this would be the discovery of more sources and their respective multiple images, but this is not guaranteed. One avenue worth exploring is the treatment of Group A as an extended source, and seeing how this changes the mass distribution, the south-east clump, and the notable presence of extraneous images in this work and in others that use $\Delta t_{DC}$ \citep{forestoribio22,liu23}. Similarly, another worthwhile study would be a line of sight analysis of the nearby foreground galaxy to see if this affects the south-east clump. Depending on its redshift, the implied mass of the clump may be even larger than our measured mass. 

In general, our mass distributions reproduce the observed images and time delays accurately, and are broadly consistent with previous reconstructions. Furthermore, we find tighter constraints on the central QSO image time delay with the inclusion of the newly measured $\Delta t_{DC}$. 

\section*{Acknowledgements}
The authors acknowledge the computational resources provided by the Minnesota Supercomputing Institute (MSI), which were critical for this study.


\section*{Data Availability}

Data generated from this article will be shared upon reasonable request to the corresponding author.

\bibliographystyle{mnras}
\bibliography{references}




\appendix


\bsp	
\label{lastpage}
\end{document}